\documentclass{aa}  

\usepackage{graphicx}
\usepackage{subfig}
\usepackage{txfonts}
\usepackage[flushleft]{threeparttable}
\usepackage{mathtools}
\usepackage{multirow}
\usepackage{makecell}
\usepackage{booktabs}
\usepackage{colortbl}
\usepackage{comment}
\usepackage{stfloats}
\usepackage{float}
\usepackage{afterpage}
\usepackage{hyperref}  
\hypersetup{colorlinks=true,linkcolor=[rgb]{0.1,0.4,1.},citecolor=[rgb]{0.1,0.4,1.},filecolor=[rgb]{0.7,0.2,0.2},urlcolor=[rgb]{0.1,0.4,1.}}

\usepackage{color}
\definecolor{blue}{rgb}{0., 0., 1}
\definecolor{lightblue}{rgb}{0.1,0.4,1.}

\usepackage{soul}
\usepackage{multicol}

\begin{document} 

\title{The accretion history of the Milky Way \\ 
\Large V. The kinematics of most globular clusters trace the merger epochs} 

\titlerunning{GC kinematics trace the merger epochs}
\authorrunning{I. Akib et al.}

\author{Istiak Akib\inst{\ref{ins1}, \ref{ins2}} \fnmsep\thanks{E-mail: \href{mailto:istiak-hossain.akib@obspm.fr}{istiak-hossain.akib@obspm.fr}}
\and François Hammer\inst{\ref{ins1}}
\and  Yanbin Yang\inst{\ref{ins1}}}

\institute{
LIRA, Observatoire de Paris, Université PSL, CNRS, Place Jules Janssen 92195, Meudon, France \label{ins1}
\and 
Center for Astronomy, Space Science and Astrophysics, Independent University, Bangladesh, Dhaka 1229, Bangladesh \label{ins2}}

\date{Received 16 February 2026 / Accepted 1 June 2026}

\abstract{Several studies have associated globular clusters (GCs) with former Galactic accretion events by comparing their positions in the energy-angular momentum ($E$-$L_z$) plane, an approach further supported by similarities in their age-metallicity relations. However, recent merger simulations suggest that GCs initially associated with the Gaia-Sausage-Enceladus (GSE) disc may have lost their orbital energy and thus may not reliably trace this accretion event. We extend this framework by considering three N-body simulations of the Milky Way-GSE merger with different initial masses, mass ratios, and gas content. In addition to GCs belonging to the GSE disc progenitor, we accounted for GCs in its halo and, in gas-rich models, a population of GCs formed during the Milky Way-GSE merger. We confirm that most GCs originating in the disc have lost a significant part of their orbital energy during repeated passages through the dense disc medium, and we conjecture that associated tidal shocks may have destroyed many of them. In contrast, GCs from the halo and GCs formed during the merger have largely retained their orbital energy, which remains comparable to that of GSE stars even up to 9 Gyr after the completion of the merger. By using a more realistic GC population and GSE modelling, we find that most GCs linked to GSE can be associated with Milky Way accretion events in the $E$-$L_z$ plane, which supports previous observational associations based on a combination of energy-angular momentum and age-metallicity relations.}


\maketitle

\section{Introduction}
The field of Galactic archaeology has progressed rapidly since the advent of {\it Gaia} and numerous spectroscopic surveys, enabling increasingly precise determinations of the age–metallicity relation (AMR). \citet{Massari2019} also introduced how globular clusters (GCs) can be associated with the accretion event debris in the Milky Way (MW)
halo as well as with in-situ formation in Galactic structures such as the bulge or disc. The orbital energy ($E_{\rm orbital}$) and angular momentum ($L_z$) of Gaia–Sausage–Enceladus (GSE) stars can be compared to those of GCs. This selection shows a well-defined track in the AMR \citep{Massari2019, Kruijssen2020, Malhan2022}, and also helps to remove interlopers such as NGC 288 \citep{Ceccarelli2025}. This approach show that a significant fraction of GCs is associated with GSE debris in the halo, in contrast to the in-situ GC population \citep{Massari2019, Malhan2022}. The above studies also helped establish the infall time-energy relation \citep{Gott1975}, which appears to be well defined in the MW \citep{Hammer2021,Hammer2023}, and enable comparisons with MW-analogues determined from cosmological simulations \citep{Hammer2024b}.

Associations of GCs with accretion events based on the ($E_{\rm orbital}$, $L_z$) plane need to be tested using merger simulations. \citet[hereafter P23]{Pagnini2023} have shown that most GCs associated with the GSE progenitor disc have lost a significant part of their orbital energy during and after the merger. This energy loss would significantly affect the accuracy of the method that associates GCs to past merger debris, as the parameter space would effectively be reduced from four (AMR, $E_{\rm orbital}$,$L_z$) to two (AMR). However, the conclusions of P23 are based only on GCs associated with the GSE disc progenitor. This contrasts with the GC population of the MW, which is dominated by GCs in the halo and bulge, with only a small fraction associated with its disc \citep{Massari2019, Malhan2022}. There are three other major reasons to support the idea that most GCs cannot be associated with galactic discs. First, GCs in a galactic disc are strongly affected by their repeated passage through the dense disc medium, leading to their rapid destruction \citep{Aguilar1985, Aguilar1988, Aguilar1993}, while GCs in the halo and bulge can survive due to spending much of their time near the apocentre in a low-density environment \citep{Hammer2023}. Second, some GCs are likely formed during the strong star-forming phase triggered by the GSE gas-rich merger \citep{DeLucia2024, Valenzuela2024}, which requires hydrodynamical N-body simulations to be modelled properly. Third, a galaxy similar to the MW may have experienced many ancient mergers \citep{Akib2025} that likely deposited GCs in the halo of the remnants.

This paper is the fifth in our series on the MW accretion history. Previous papers established the energy–accretion time history \citep{Hammer2023}, investigated the dynamical evolution of GCs and dwarf galaxies \citep{Hammer2024a, Wang2024}, and reported the discovery of young stars in dwarf spheroidal galaxies, providing further support for their late infall into the MW halo \citep{Yang2024}. Here, we extend the method of P23 by recovering the full population of GCs expected in galaxies. In Sect. \ref{sec: Pagnini23}, we discuss how we re-ran one of the models presented in P23 with an additional GC population in the GSE halo, and we analyse both the spatial distribution and the distribution in the $E-L_z$ plane. In Sect. \ref{sec: MWlike}, we analyse the distribution of GC populations in two more realistic hydrodynamical merger models that include gas and star formation, thereby allowing us to account for GCs formed during the merger as a third GC population. Finally, in Sect. \ref{sec: Discussion} we discuss the impact and limitations of this study.

\section{Introducing globular clusters in the GSE halo}
\label{sec: Pagnini23}
We re-ran the N-body merger model MWsat{\_}n1{\_}$\Phi$60 from P23 but included an additional GC population in the GSE halo. The differences between the two simulations are as follows: (a) The stellar mass of P23 was modelled by three Miyamoto-Nagai discs representing the Galactic thin disc, the young thick disc, and the old thick disc \citep{Haywood2013, diMatteo2016}, whereas we adopted a single exponential stellar disc with a mass and scale length that results in a similar mass distribution. (b) The dark matter (DM) halo in P23 was modelled with a Plummer profile, while for this work we used a Hernquist DM halo whose scale length provides a similar mass distribution. (c)  In P23 the TreeSPH code was used \citep{Semelin2002}, whereas we used GIZMO \citep{Hopkins2015} with adaptive time stepping as well as different softening and tolerance schemes.\footnote{We note that P23 adopted a fixed time step of $\Delta t = 2.5 \times 10^{5}$ yr and Plummer gravitational softening within TreeSPH. GIZMO employs an adaptive time-stepping scheme (with a typical value of $\Delta t = 10^{4}$ yr in our simulations) and a modern spline softening kernel. These choices provide significantly improved accuracy in following the system’s dynamical evolution.} The parameters of the components of our Model 1 are shown in Table \ref{tab:MW_models}. A comparison between the stellar and DM density profiles of Model 1 in this work and MWsat{\_}n1{\_}$\Phi$60 from P23 is presented in Appendix \ref{app:Model 1}. The ratio of total mass between the secondary and primary galaxy ($\mu_{\rm gal}$) in this model is 1:10.

\begin{table*}
\centering
\caption{Initial MW and GSE components and orbital configurations.}
\label{tab:MW_models}

\begin{threeparttable}
\small
\begin{tabular}{cccccccccccccccccc}
\toprule
Merger &
Comp- &
\multicolumn{3}{c}{MW} &
\multicolumn{3}{c}{GSE} &
\multicolumn{6}{c}{GSE Initial 6D Coordinates} &
\multicolumn{4}{c}{Orbital Angles} \\
\cmidrule(lr){3-5}
\cmidrule(lr){6-8}
\cmidrule(lr){9-14}
\cmidrule(lr){15-18}
 Model & onents & $M$ & $a$ & $h$
 & $M$ & $a$ & $h$
 & $x$ & $y$ & $z$ & $v_x$ & $v_y$ & $v_z$
 & $i_1$ & $w_1$ & $i_2$ & $w_2$ \\
\midrule

\multirow{2}{*}{\makecell{Model 1\\$\mu_{\rm gal}$ = 1:10}}
& Star
& 8.35 & 3.35 & 0.425
& 0.85 & 1.1  & 0.14
& \multirow{2}{*}{50}
& \multirow{2}{*}{0}
& \multirow{2}{*}{-86.6}
& \multirow{2}{*}{-103}
& \multirow{2}{*}{42}
& \multirow{2}{*}{178}
& \multirow{2}{*}{0}
& \multirow{2}{*}{60}
& \multirow{2}{*}{0}
& \multirow{2}{*}{240} \\
& DM
& 36.8 & 15   & --
& 3.68 & 2.5  & --
&      &      &       &       &       & 
&      &      &      &      \\
\midrule

\multirow{3}{*}{\makecell{Model 2\\$\mu_{\rm gal}$ = 1:3}}
& Star
& 0.85 & 1.6  & 0.3
& 0.15 & 0.9  & 0.2
& \multirow{3}{*}{80}
& \multirow{3}{*}{60}
& \multirow{3}{*}{0}
& \multirow{3}{*}{80}
& \multirow{3}{*}{160}
& \multirow{3}{*}{0}
& \multirow{3}{*}{70}
& \multirow{3}{*}{30}
& \multirow{3}{*}{-110}
& \multirow{3}{*}{30} \\
& Gas
& 7.55 & 12.5 & 2.5
& 2.85 & 7.0  & 1.4
&      &      &     &     &     &
&      &      &      &      \\
& DM
& 20.5 & 5.6  & --
& 6.8  & 3.8  & --
&      &      &     &     &     &
&      &      &      &      \\
\midrule

\multirow{3}{*}{\makecell{Model 3\\{$\mu_{\rm gal}$} = 1:3}}

& Star
& 2.0  & 2.64 & 0.156
& 0.28 & 1.29 & 0.083
& \multirow{3}{*}{-63}
& \multirow{3}{*}{-76}
& \multirow{3}{*}{0}
& \multirow{3}{*}{95}
& \multirow{3}{*}{255}
& \multirow{3}{*}{0}
& \multirow{3}{*}{110}
& \multirow{3}{*}{30}
& \multirow{3}{*}{-110}
& \multirow{3}{*}{30} \\
& Gas
& 6.1  & 21.2 & 0.31
& 2.5  & 11.4 & 0.149
&      &      &     &     &     &
&      &      &     &      \\
& DM
& 75   & 17.5 & --
& 25   & 10   & --
&      &      &     &     &     &
&      &      &     &      \\
\bottomrule
\end{tabular}
\begin{tablenotes}[flushleft]
\footnotesize
\item \textit{Notes.} Masses ($M$) are given in units of $10^{10}$ M$_\odot$; scale lengths ($a$) and scale heights ($h$) are given in kiloparsecs. The six-dimensional phase-space coordinates $(x,y,z,v_x,v_y,v_z)$ are given in the Galactocentric frame, with distances in kiloparsecs and velocities in km s$^{-1}$. Orbital angles are given in degrees. For Model 1, the masses of individual stellar and DM particles are $1.4 \times 10^{5}$ M$_\odot$ and $5.6 \times 10^{5}$ M$_\odot$, respectively. For Models 2 and 3, baryonic (star or gas) and DM particles have masses of $0.6 \times 10^{5}$ M$_\odot$ and $2.3 \times 10^{5}$ M$_\odot$, respectively.
\end{tablenotes}
\end{threeparttable}
\end{table*}

Similar to P23, the first population of GC particles was mimicked by randomly sampling stellar particles from the initial GSE disc, and they are referred to as 'Disc GCs'. For the second population, consisting of GCs in the halo, we randomly sampled DM halo particles located at distances between five and ten times the disc scale length from the GSE centre. These are hereafter referred to as 'Halo GCs'. Since the DM particles are initially in dynamical equilibrium, and Halo GCs and stars evolve primarily under gravity without significant dynamical friction (given the low baryonic density in these regions), the DM particles can serve as suitable proxies for GCs in the halo under a point-mass approximation. The adopted distance range for selecting Halo GCs is motivated by the observed distribution of GCs in the MW halo \citep{Hammer2023}. The majority of MW GCs are found within $\sim 25$ kpc, corresponding to approximately ten times the Galactic disc scale length. Although GCs are observed at distances up to $\sim 150$ kpc, such GCs in the outer-halo are only weakly bound to the central potential and are therefore less susceptible to significant energy loss. Consequently, restricting our sample to the chosen radial range provides a representative and dynamically relevant test population for our analysis.

\begin{figure*}[p]
\centering
    \includegraphics[width=1.85\columnwidth]{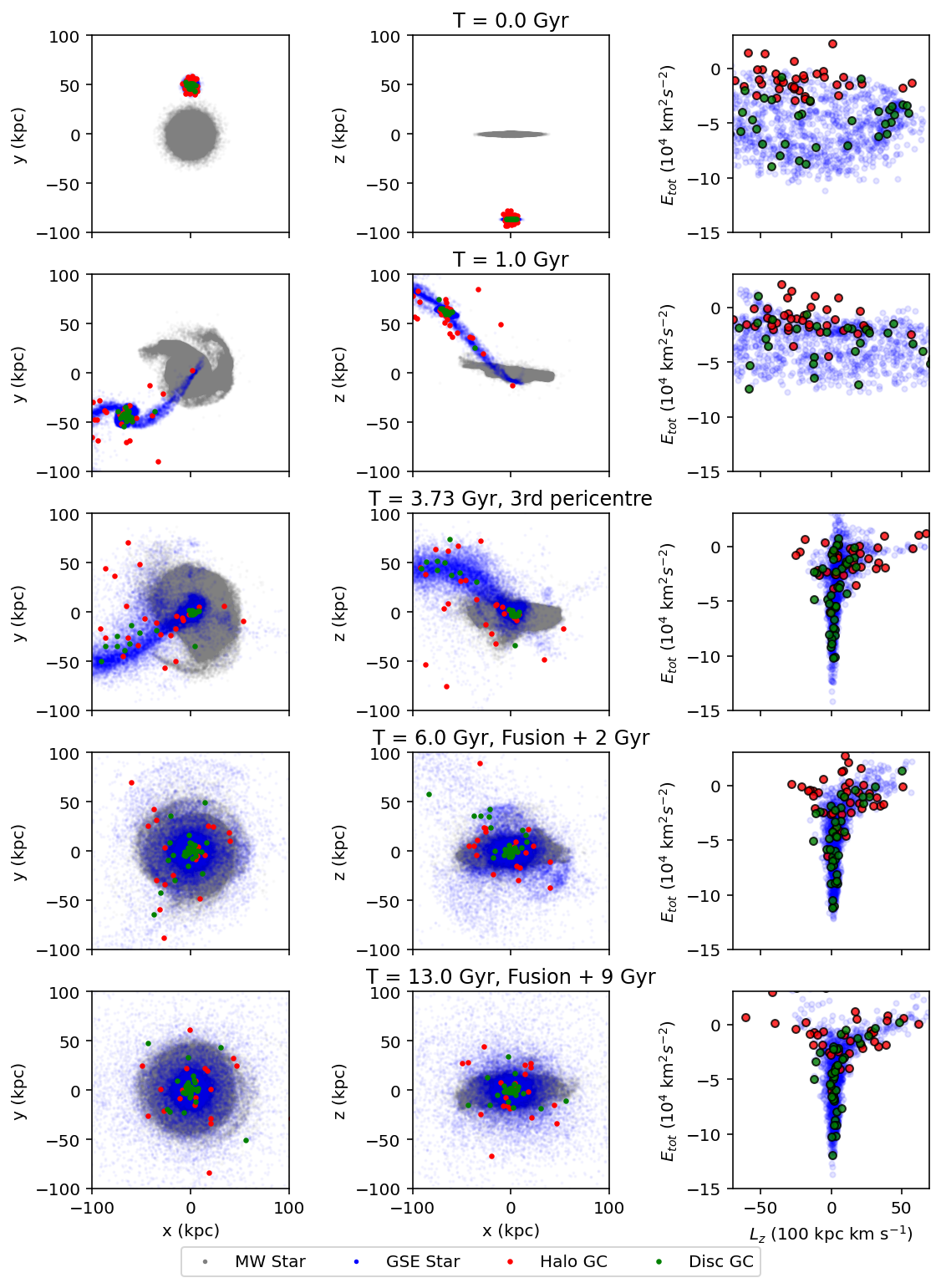}

 \caption[Time evolution]{Time evolution of the merger process for Model 1. Left, middle columns: Projections of the simulated GC positions onto the $x-y$ and $x-z$ planes. The $z-$axis is aligned with the direction of angular momentum at each epoch. The MW stars are shown in grey, and GSE stars are in blue. The GCs originally associated with the GSE disc are shown in green, and those originating in the GSE halo are shown in red. Right column: Distributions of GSE stars and GCs in the $E-L_z$ space at the corresponding times.}
 \label{GP f1}
\end{figure*}

Figure \ref{GP f1} shows the $x-y$, $x-z$, and $E-L_z$ distributions as a function of time for Model 1, and it is analogous to Fig. 1 of P23. The fusion epoch of Model 1 ($\sim$ 4 Gyr) is different from that of P23 ($\sim$ 3.2 Gyr), possibly due to differences in the simulation code, gravitational softening, and time-stepping schemes. We defined the fusion epoch as the time at which the distance between the cores of the two galaxies became consistently less than 1 kpc. We associated the third row of panels in the figure with the third pericentric passage at 3.73 Gyr and the fourth row with 2 Gyr after fusion. Since the MW-GSE merger is estimated to have occurred about 9-10 Gyr ago \citep{Belokurov2018, Haywood2018, Helmi2018}, we also included a fifth row corresponding to 9 Gyr after fusion, which would be representative of the present-day MW. Figure \ref{GP f1} reproduces the key features seen in the Fig. 1 of P23: Most Disc GCs lose energy and become more tightly bound in the MW due to strong gravitational torques and tidal effects during the merger. The $x-z$ panels show that these GCs predominantly settle into the disc, while all GCs associated with the GSE have eccentricities higher than 0.6 \citep{Hammer2023}. Hence, the Disc GCs that lose energy in this manner and end up in the disc would not be representative of the GCs that are associated with the GSE in observational studies. However, the Halo GCs appear to retain their high energy because they are far less affected by the merger, due to their initial locations in the halo. We note that many of them also exhibit retrograde orbits, similar to some GSE stars, even 9 Gyr after the fusion.

\section{Globular clusters in hydrodynamical simulations}
\label{sec: MWlike}
Our analysis of the N-body 1:10 minor merger model from P23 shows that GCs originally in the GSE halo can retain their energy. However, the models of P23 may not be sufficient to fully describe the GC population of the MW for several reasons. First, the models in P23 are not designed to reproduce the currently observed properties of the MW, such as the timing of the MW-GSE merger or the mass distribution of the MW. Second, P23 assumes a minor merger with a mass ratio of 1:10, whereas there is evidence from observations and modelling that the ratio could be as high as 1:3 \citep{Helmi2018, Vincenzo2019, Naidu2021}. A 1:3 merger is significantly more energetic than a 1:10 merger; therefore, it is possible that Halo GCs would be more strongly affected than in the minor-merger case. Third, P23 does not include gas and therefore lacks star formation. The MW-GSE merger occurred at $z$ > 1, when galaxies were generally gas-rich. The star formation history of the MW suggests that around 80\% of its stellar mass formed after the first passage of the MW-GSE merger \citep{Haywood2016}. For these reasons, we tested the kinematics of GCs using merger simulations with higher mass ratios and including star formation. We analysed two models representing gas-rich major mergers ($\mu_{\rm gal}=1{:}3$), inspired by the simulations of \citet{Sauvaget2018}. The baryonic scale lengths of the primary and secondary galaxies in \citet{Sauvaget2018} are based on observations of galaxies in the early Universe. We considered Model 2, a low-mass Milky Way model whose rotation curve is consistent with \citet{Jiao2023}, and Model 3, a high-mass Milky Way model whose rotation curve is similar to that of \citet{Eilers2019}. The rotation curves from the models and observations are shown in Appendix \ref{app:rc}, and the initial conditions of these models are listed in Table \ref{tab:MW_models}. To construct these models, we started from the set initial conditions of \citet{Sauvaget2018} and adjusted them to reproduce the desired mass and the expected GSE signature. Among the tested configurations, inclined prograde-retrograde orbits best reproduce the GSE signature and other properties of the MW. However, identifying this signature from kinematics alone is non-trivial. The primary challenge arises from contamination by MW stars with similar dynamical properties, as highlighted by \citet{Carrillo2024}. Consequently, the GSE signature alone is insufficient to uniquely constrain the initial orbital parameters. For Model 3, we adopted an inclined retrograde-prograde orbit to explore its impact on GC kinematics.

Since Models 2 and 3 include gas and star formation, we were able to define a third sample, which we refer to as 'Merger GCs'. The merger of gas-rich galaxies is accompanied by a period of enhanced star formation, which declines after the fusion is complete (see Appendix \ref{app:sfh}). This process can lead to the formation of star clusters \citep{Valenzuela2024}, some of which may survive if they remain outside dense regions. To mimic this population in our simulations, we selected stellar particles formed from GSE gas during the star formation epoch associated with the merger. From this population, we further selected those located in the tidal tails during the fusion of the cores of the two galaxies, thereby giving these candidate GCs the highest likelihood of survival. At this epoch, a minimum distance threshold of 35 kpc for Model 2 and 25 kpc for Model 3 was empirically adopted to isolate these tidal tail particles. This GC population is discussed further in Section \ref{sec: Discussion}.

\begin{figure*}[p]
\centering
    \includegraphics[width=1.85\columnwidth]{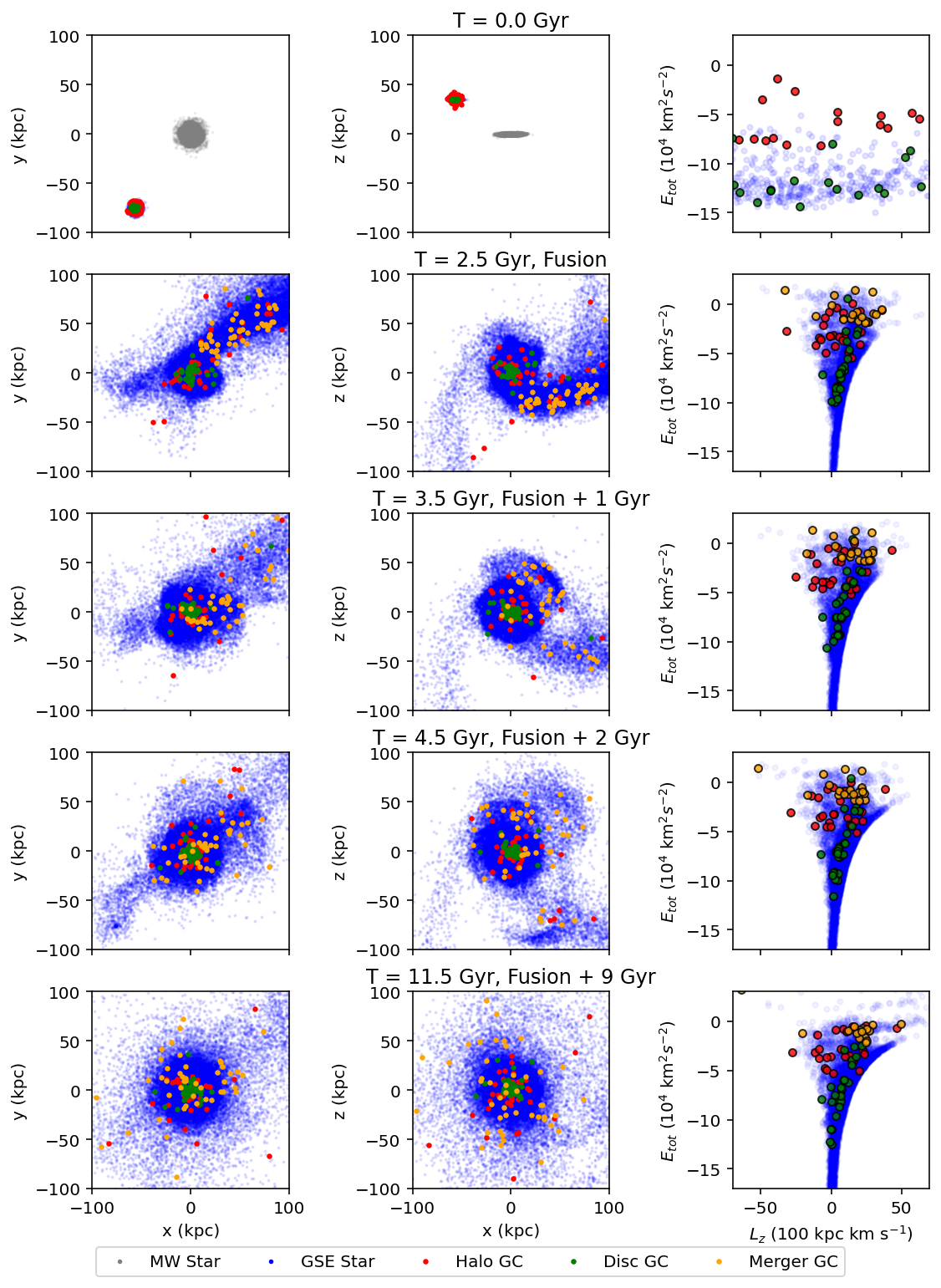}

 \caption[Time evolution of Model 2]{Time evolution of the merger process for Model 2. Left, middle columns: Projections of the simulated GC positions onto the $x-y$ and $x-z$ planes. The $z-$axis is aligned with the direction of angular momentum at each epoch. The MW stars are shown in grey, and GSE stars are in blue. The GCs originally associated with the GSE disc are shown in green, and those originating in the GSE halo are shown in red. The Merger GCs formed from GSE gas in the tidal tail during the merger and are shown in orange. Right column: Distributions of GSE stars and GCs in the $E-L_z$ space at the corresponding times.}
 \label{LM f1}
\end{figure*}

Figure \ref{LM f1} shows the evolution in both spatial distribution and in an $E$ versus $L_z$ plot for Model 2. A similar plot for Model 3 is presented in Appendix \ref{app:Model 3}. For both models, the behaviour of the Disc GCs and Halo GCs is consistent with that of Model 1; that is, the GCs initially located in the GSE disc lose their orbital energy over time and become more tightly bound, whereas those originating in the GSE halo largely retain their energy. We further found that Merger GCs also preserve their high energies, which are comparable to that of Halo GCs. The spatial distribution of GCs is significantly more concentrated in Model 3 than in Model 2, which can be attributed to Model 3 being approximately three times more massive. Furthermore, the fraction of GCs and stars on retrograde orbits is lower in Model 3 than in Model 2. This difference likely arises from the initial orbital configuration of GSE, which is prograde in Model 3 but retrograde in Model 2, and may also be influenced by the higher system mass in Model 3.

\section{Discussion and conclusion}
\label{sec: Discussion}
\begin{figure*}[ht!]
\centering
    \includegraphics[width=2\columnwidth]{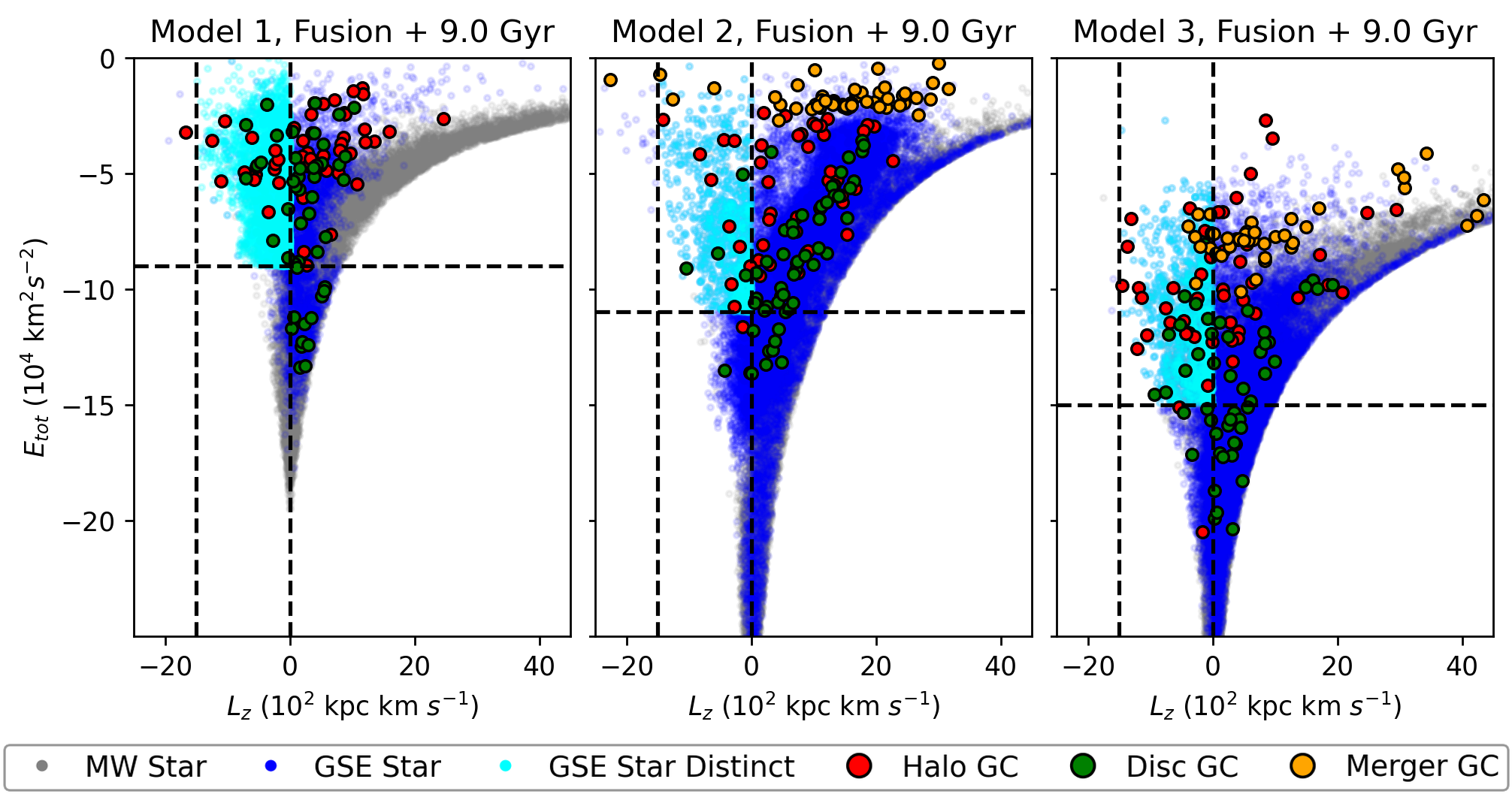}

 \caption[E-Lz] {Distributions of stars and GCs of the three models in the $E-L_z$ space. The panels show MW stars (grey), GSE stars (blue), and the subset of GSE stars that can be distinguished from the MW population (cyan). Different populations of GCs associated with GSE are shown: Halo GCs (red), Disc GCs (green), and Merger GCs (orange).}
 \label{E-Lz plot}
\end{figure*}

Our simulations show that a majority of the GCs can keep their high orbital energy if they were not initially located in the GSE disc. Figure \ref{E-Lz plot} shows the $E$ versus $L_z$ distributions for the three models at 9 Gyr after fusion. To better match the observational constraints, we restricted the stellar and GSE GC samples to a maximum galactocentric distance of 30 kpc. In Model 1, stars have higher average energy than in Model 2, despite Model 1 having a higher total mass. This is because the scale length of the DM halo in Model 1 is much larger than that in Model 2, even after accounting for the mass difference. The identification of GSE debris from observation in the $E-L_z$ plane is complicated by the contamination from MW stars. Although GSE stars span a wide range in the $E-L_z$ space, GSE debris have been identified as stars in retrograde orbits and with high orbital energy, which makes them easier to distinguish from MW stars \citep{Helmi2018, Gaia2023, Carrillo2024}. An energy cut was applied at the point where GSE stars begin to deviate from the trend of the disc, and this limit was determined by visual inspection for each model. For Models 2 and 3, a significant fraction of the GSE stars align with MW stars. Once the merger is complete, the disc is formed from stellar and gaseous material from both galaxies. Since stars are treated as point masses and are affected only by gravity, multiple repeated interactions with the dense medium are required for them to lose sufficient energy and align with the disc. As a result, stellar particles originally belonging to GSE do not acquire sufficiently high angular momentum to participate in disc rotation (see the GSE stars of Model 1 in Figure \ref{E-Lz plot}). In contrast, gas can dissipate energy efficiently through hydrodynamical interactions and therefore settle into the newly formed disc. Stars formed from this gas inherit the kinematics of the disc. Consequently, in Models 2 and 3, we observed some GSE-associated stars aligning with the disc in Figure \ref{E-Lz plot}. This feature is absent in Model 1 since the model does not include gas or star formation.

Across all models, Halo GCs exhibit higher energies and a wider range of $L_z$ than Disc GCs, most of which have lost a significant fraction of their energy. The Merger GCs occupy even higher energy orbits than the Halo GCs due to their later accretion into the MW halo as material returning from tidal tails. However, their angular momentum is more tightly clustered compared to that of the Halo GCs. This is because they originate in infalling tidal tails and therefore retain a preferential direction of rotation, unlike the more randomly oriented Halo GCs. The Halo GCs and Merger GCs span a much wider range in $L_z$ than has been suggested in the literature for GSE \citep{Massari2019, Malhan2022}. This discrepancy could be linked to the initial orbital configuration of the mergers. Under a different orbital setup, it may be possible to obtain a $L_z$ distribution for GCs that is, on average, more retrograde and spans a more limited range, as identified in observational studies. Observationally, GCs spanning a wide range of angular momentum ($L_z$) are found at energies comparable to those associated with the GSE. These GCs are typically attributed to distinct accretion events, such as Sequoia and Arjuna for more retrograde populations, and the Helmi stream, LMS-1/Wukong for more prograde populations \citep{Massari2019, Malhan2022}. However, the inferred epochs of these events significantly overlap with that of the GSE merger \citep{Kruijssen2020}. Furthermore, the metallicities of GCs associated with these events also overlap with those of GSE-associated GCs \citep{Malhan2022}. Since GCs serve as key tracers for identifying and dating accretion episodes, this raises the possibility that the GSE merger alone may be sufficient to explain these GC populations and the associated inferred accretion events.

In all three models, we found that most of the Disc GCs lose their energy, confirming the findings of \cite{Pagnini2023}. However, if they lose their energy due to tidal shocks during repeated passages through a dense disc medium, the latter would likely destroy them \citep{Aguilar1985, Aguilar1988, Aguilar1993}. Since we represent the GCs as point masses (both stellar and DM particles), these disruptions cannot be captured in our simulations. Representing GCs as clumps of several stellar particles, such as in \cite{Li2004}, would require an N-body hydrodynamical simulation with hundreds of millions of particles since GCs are much less massive than the MW. In such a framework, the distribution of GCs would likely be more consistent with observations, assuming that most Disc GCs that lose orbital energy are subsequently destroyed.

Our introduction of a third population of GCs associated with the merger was motivated by both observations and theory. \cite{Valenzuela2024} provide observational evidence that gas-rich (wet) mergers are associated with bursts of cluster formation in the form of strong HII regions that may become progenitors of some GCs on the condition that they can escape the dense disc medium, for example, after being captured into a tidal tail (see Tsakonas et al. 2026, in preparation, for the description of this channel for the GC formation based on deep observations of M31.) Such a scenario is consistent with multimodal age distributions in the GC systems of the MW and nearby galaxies that correlate with the timing of past mergers. \cite{Elmegreen2018} showed that GC formation requires exceeding two key thresholds: high gas surface density and a high star formation rate. Both of these conditions can naturally be overcome during gas-rich mergers, as shown in the high-resolution simulations in \cite{Li2004}. Our Merger GCs, sampled from the stellar particles formed during the starburst phase, trace these high density, high star formation rate environments. Furthermore, \cite{DeLucia2024}, using cosmological simulations, proposed a two-phase paradigm in which GCs may form in galactic discs but preferentially survive if they are displaced into the halo during mergers, where they are less susceptible to destructive tidal forces. Selecting newly formed stellar particles that are transported into tidal tails during the merger therefore provides a physically motivated proxy for clusters that both form during merger-driven starbursts and have an enhanced likelihood of long-term survival. 

We conclude that realistic modelling of GC evolution during mergers such as the GSE supports their association with merger events through the $E$–$L_z$ and AMR. This is consistent with previous studies \citep{Massari2019, Kruijssen2020, Malhan2022} and more recently with \cite{Massari2026}.

\begin{acknowledgements}
We would like to thank the referee for their insightful comments and suggestions, which greatly strengthened the overall manuscript. We thank Gary Mamon and Piercarlo Bonifacio for their helpful comments on the manuscript. Simulations in this work were performed at the High-performance calculation (HPC) resources MesoPSL financed by the project Equip@Meso (reference ANR-10-EQPX- 29-01) of the program “Investissements d’Avenir” supervised by the ‘Agence Nationale de la Recherche’.
\end{acknowledgements}
\clearpage
\bibliographystyle{aa}
\bibliography{ref}

\begin{appendix}
\section{Density distributions of Model 1 and MWsat{\_}n1{\_}$\Phi$60 components}
\label{app:Model 1}
\begin{figure}[h!]
\centering
    \includegraphics[width=\columnwidth]{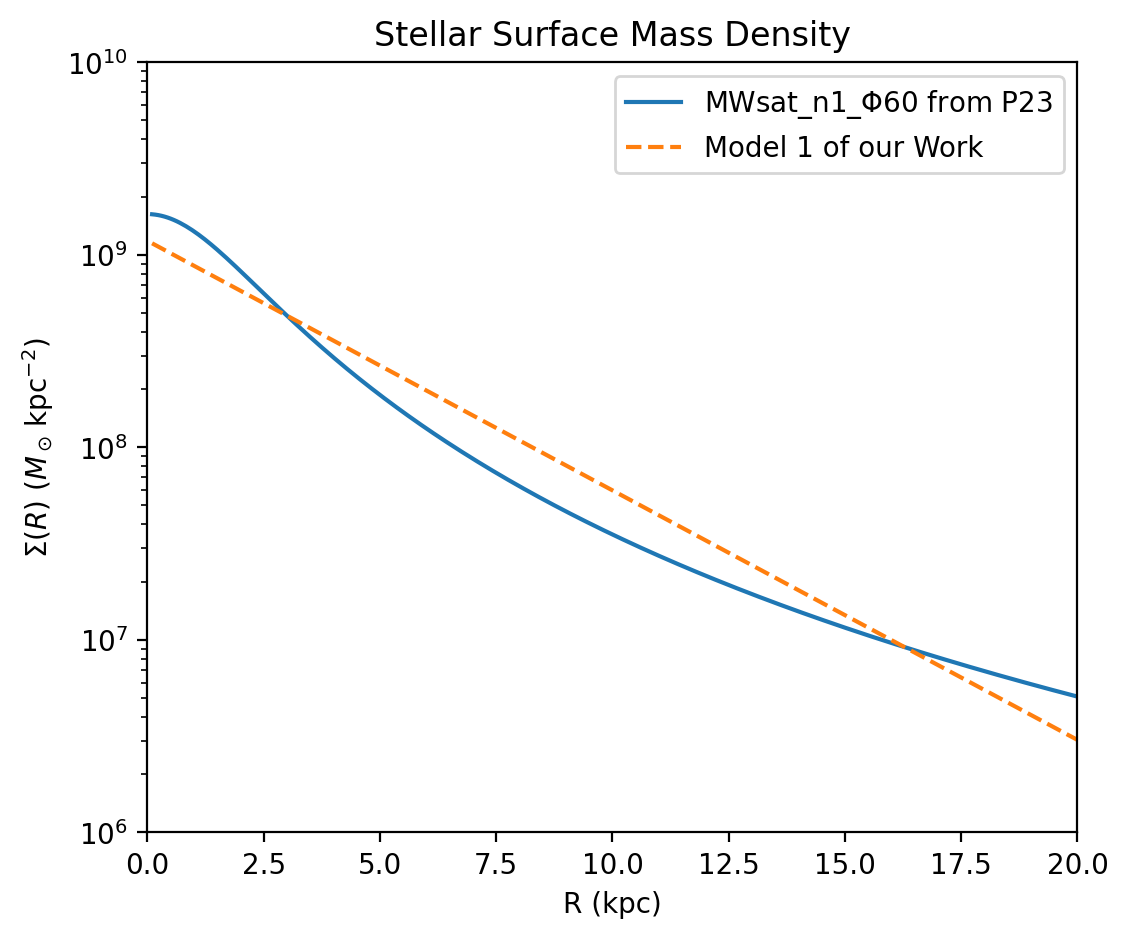}

 \caption[Stellar surface mass density]{Stellar surface mass density profiles of MWsat{\_}n1{\_}$\Phi$60 from P23 and Model 1. The stellar component of MWsat{\_}n1{\_}$\Phi$60 is a combination of three Miyamoto-Nagai discs, whereas that of Model 1 consists of a single exponential disc. }
 \label{M1 disc}
\end{figure}

\begin{figure}[h!]
\centering
    \includegraphics[width=\columnwidth]{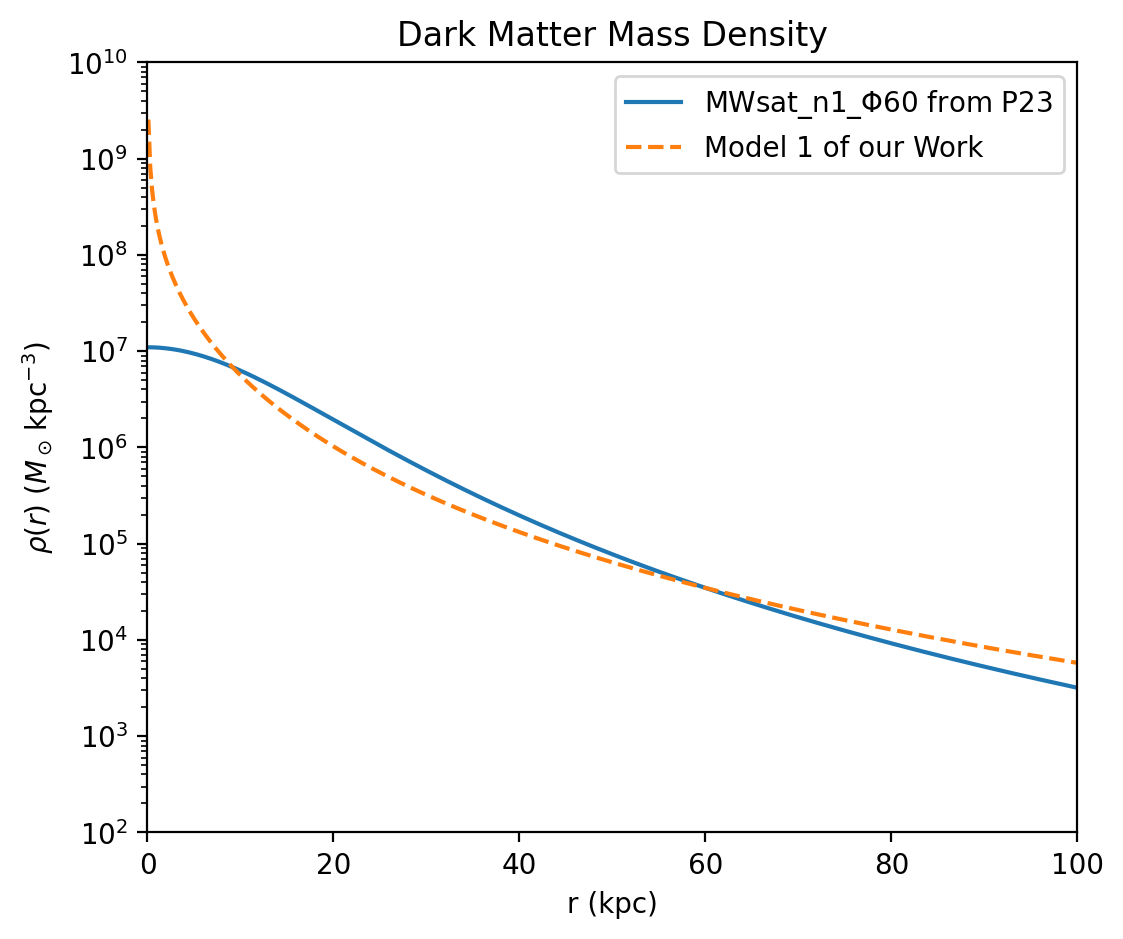}

 \caption[DM density]{Dark matter density profile of MWsat{\_}n1{\_}$\Phi$60 from P23 and Model 1. The DM halo of MWsat{\_}n1{\_}$\Phi$60 follows a Plummer profile, whereas that of Model 1 follows a Hernquist profile.}
 \label{M1 DM}
\end{figure}

\newpage

\section{Properties: Model 2 and Model 3}
\label{app:Model 2, 3}
\subsection{Rotation curve}
\label{app:rc}
\begin{figure}[h!]
\centering
    \includegraphics[width=\columnwidth]{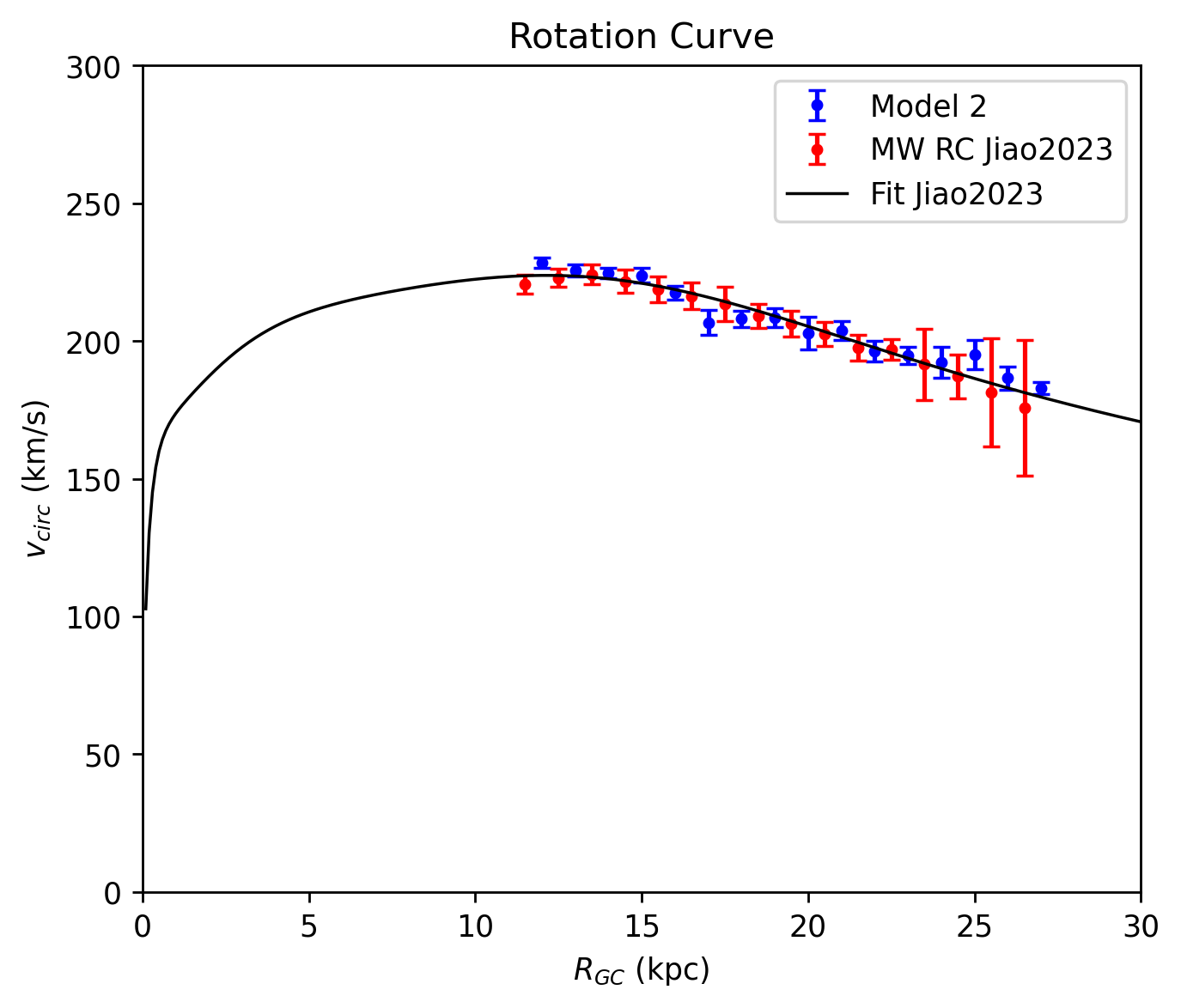}

 \caption[Rotation Curve]{Rotation curve for Model 2 (blue) compared with the Gaia DR3 rotation curve derived by \citet{Jiao2023} (red) and their best-fit model (black).}
 \label{B2 rc}
\end{figure}

\begin{figure}[h!]
\centering
    \includegraphics[width=\columnwidth]{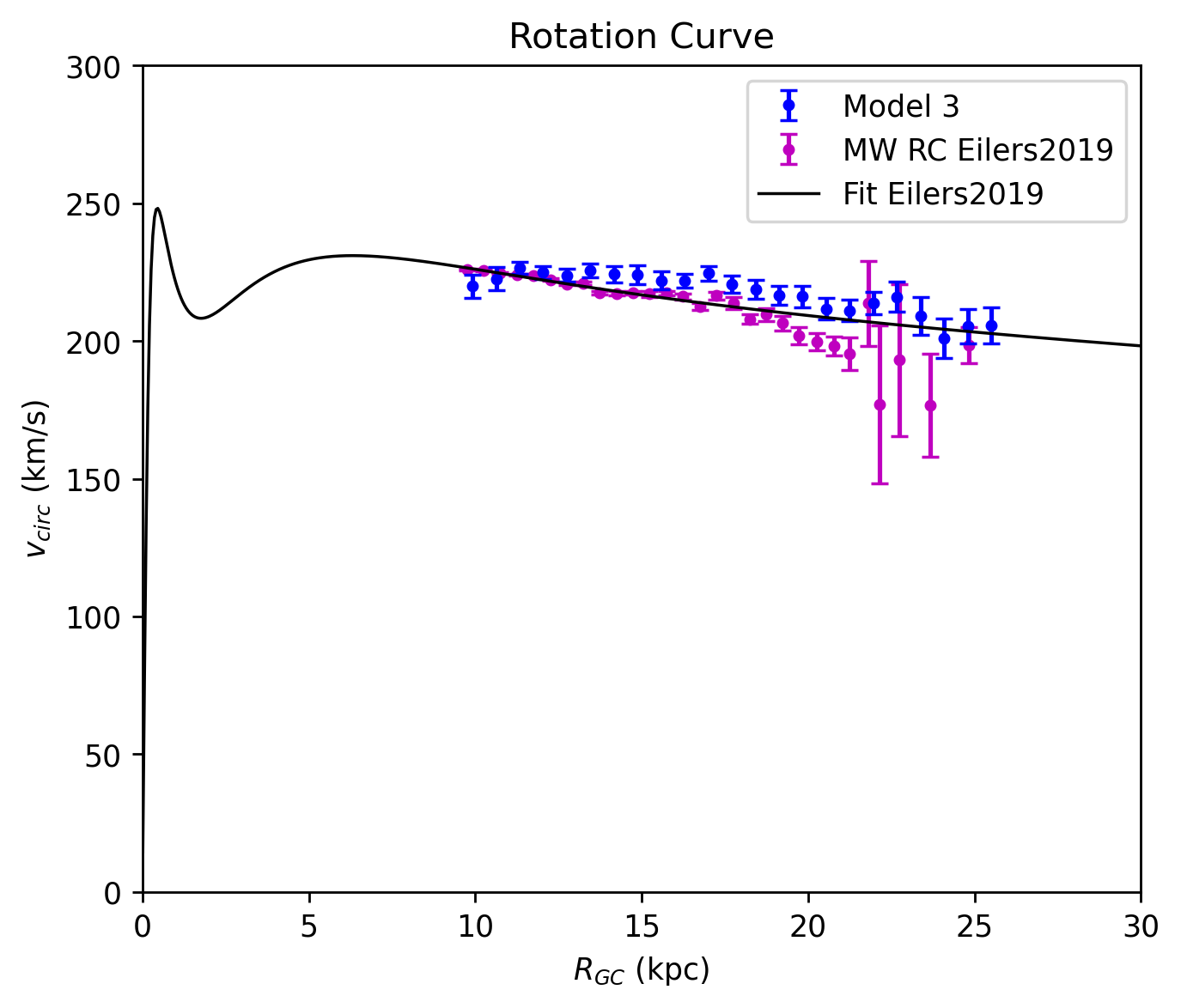}

 \caption[Rotation Curve]{Rotation curve for Model 2 (blue) compared with Gaia DR2 rotation curve derived by \cite{Eilers2019} (magenta) and their best-fit model (black)}
 \label{B3 rc}
\end{figure}

\newpage
\subsection{Star formation history}
\label{app:sfh}
\begin{figure}[h!]
\centering
    \includegraphics[width=\columnwidth]{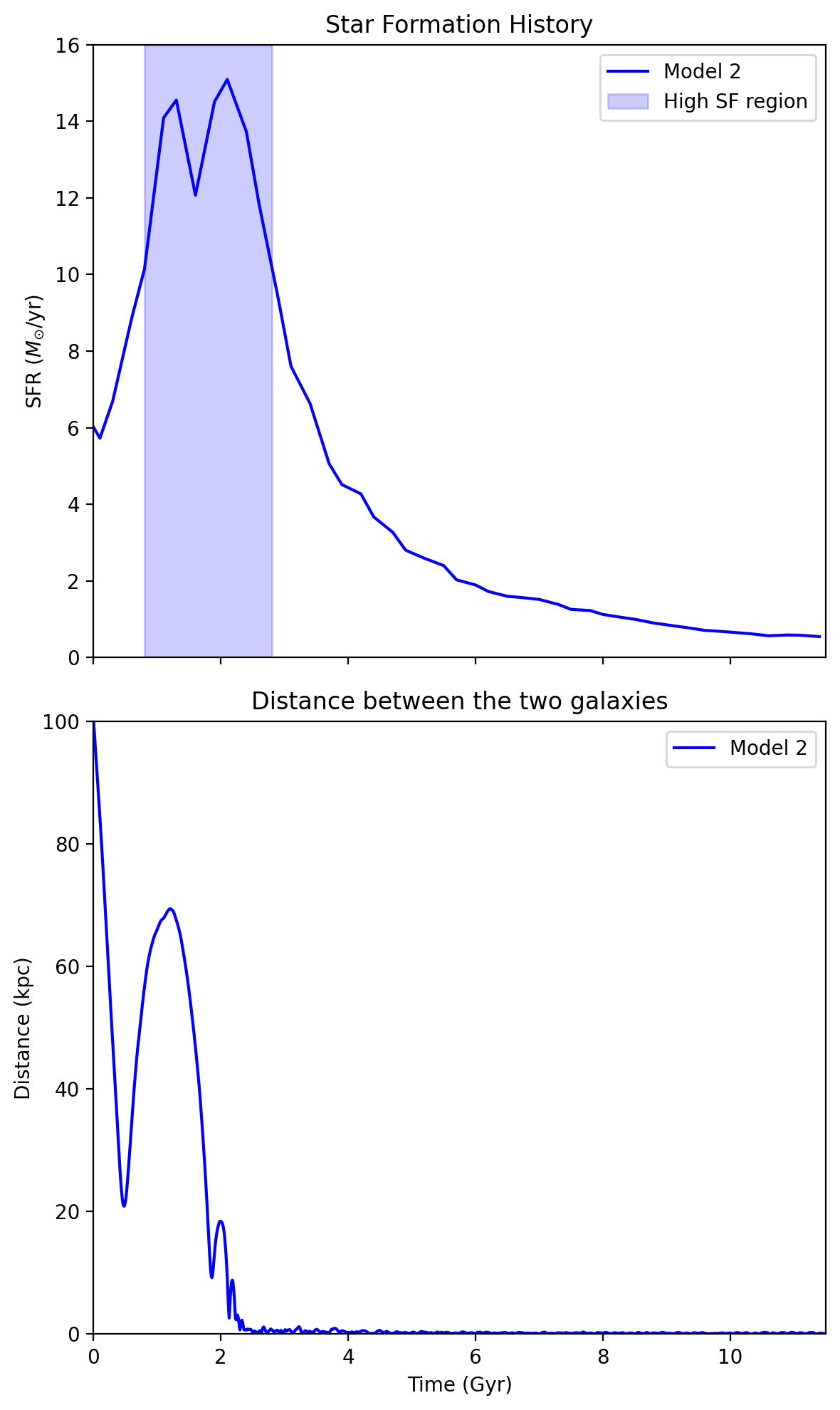}

 \caption[Star formation history]{Top panel: Star formation rate as a function of time for Model 2. The shaded region (spanning approximately 2 Gyr) highlights the enhanced star formation associated with the merger. The Merger GCs correspond to stellar particles formed during this epoch. Bottom panel: Distance between the cores of the primary (MW) and secondary (GSE) galaxies as a function of time.}
 \label{B2 sfr}
\end{figure}

\begin{figure}[h!]
\centering
    \includegraphics[width=\columnwidth]{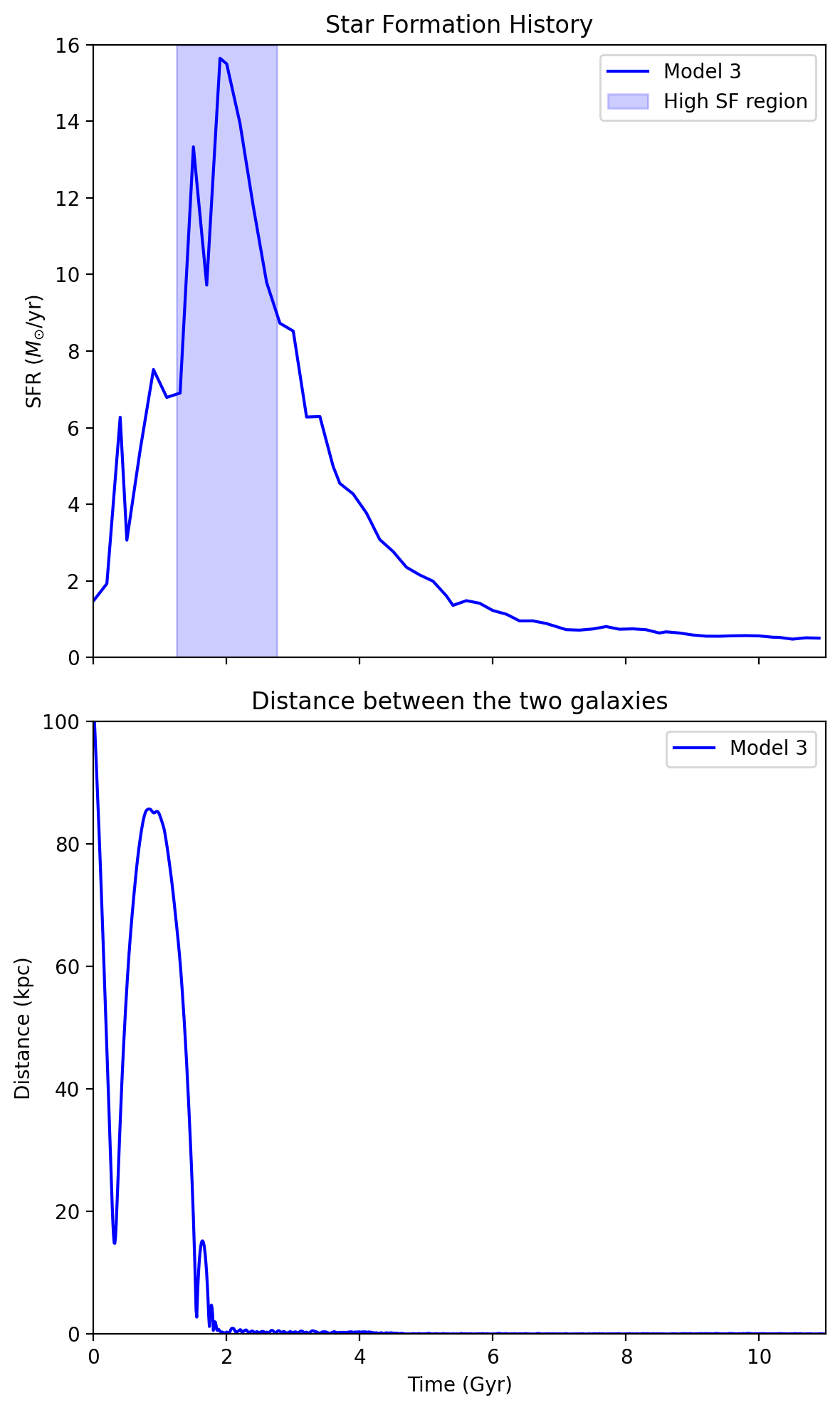}

 \caption[Star formation history]{Top panel: Star formation rate as a function of time for Model 3. The shaded region (spanning approximately 1.5 Gyr) highlights the enhanced star formation associated with the merger. The Merger GCs correspond to stellar particles formed during this epoch. Bottom panel:  Distance between the cores of the primary (MW) and secondary (GSE) galaxies as a function of time.}
 \label{B3 sfr}
\end{figure}

\newpage
\onecolumn
\section{Merger over time for Model 3} 
\label{app:Model 3}
\begin{figure*}[hb!]
\centering
    \includegraphics[width=.85\columnwidth]{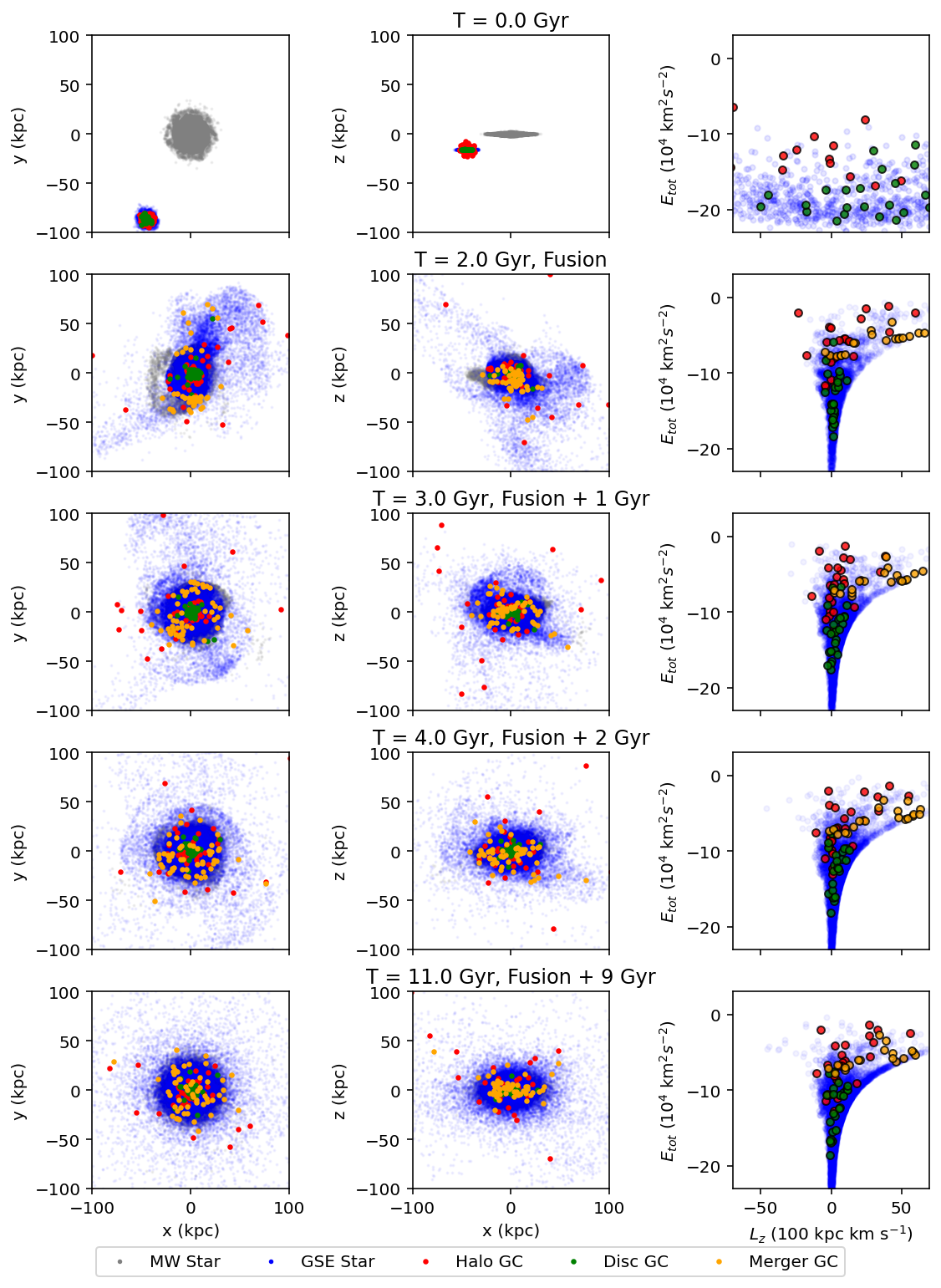}

 \caption[Time evolution of Model 3]{Time evolution of the merger process for Model 3. Left, middle columns: Projections of the simulated GC positions onto the $x-y$ and $x-z$ planes. The $z-$axis is aligned with the direction of angular momentum at each epoch. The MW stars are shown in grey, and GSE stars are in blue. The GCs originally associated with the GSE disc are shown in green, and those originating in the GSE halo are shown in red. The Merger GCs formed from GSE gas in the tidal tail during the merger and are shown in orange. Right column: Distributions of GSE stars and GCs in the $E-L_z$ space at the corresponding times.}
 \label{HM f1}
\end{figure*}
\end{appendix}

\end{document}